\documentstyle[floats,prl,aps,epsf]{revtex}
\draft 

\begin{document}

\twocolumn[\hsize\textwidth\columnwidth\hsize\csname
@twocolumnfalse\endcsname

\title{Binary Neutron Stars in General Relativity: 
	Quasi-Equilibrium Models}

\author{T.~W.~Baumgarte$^1$, G.~B.~Cook$^2$, M.~A.~Scheel$^2$,
	S.~L.~Shapiro$^{1,3}$ and S.~A.~Teukolsky$^{2,4}$}

\address{$^1$Department of Physics, University of Illinois at
        Urbana-Champaign, Urbana, Il 61801}
\address{$^2$Center for Radiophysics and Space Research, Cornell University,
        Ithaca, NY 14853}
\address{$^3$Department of Astronomy and NCSA, University of Illinois at
        Urbana-Champaign, Urbana, Il 61801}
\address{$^4$Departments of Physics and Astronomy, Cornell University,
        Ithaca, NY 14853}
\maketitle

\begin{abstract}
We perform fully relativistic calculations of binary neutron stars in
quasi-equilibrium circular orbits.  We integrate Einstein's equations
together with the relativistic equation of hydrostatic equilibrium to
solve the initial value problem for equal-mass binaries of arbitrary
separation. We construct sequences of constant rest mass and identify
the innermost stable circular orbit and its angular velocity.  We find
that the quasi-equilibrium maximum allowed mass of a neutron star in a
close binary is slightly larger than in isolation.
\end{abstract}


\vskip2pc]

The two-body problem is one of the outstanding, unsolved problems in
classical general relativity.  And yet, neutron star binary systems
are known to exist, even within our own galaxy~\cite{tamt93}.  For
some of these systems (including PSR B1913+16, B1534+12 and B2127-11C)
general relativistic orbital effects have been measured to high
precision~\cite{tw89}.  Binary neutron stars are among the most
promising sources for gravitational wave detectors now under
construction, like LIGO, VIRGO and GEO. This has triggered an intense
theoretical effort to predict the gravitational wave form emitted
during the inspiral and coalescence of the two stars.

Much of the work on binary neutron stars has been performed within the
framework of Newtonian hydrodynamics~\cite{newton}.  Many
investigators have also studied the problem in post-Newtonian (PN)
theory. As long as the PN stars are well separated, they can be
approximated by point sources~\cite{postnewton}, but for close
binaries, hydrodynamical effects must also be taken into
account~\cite{s96,on96,l96,lw96,lrs97}.

Fully general relativistic treatments of the problem are complicated
by the nonlinearity of Einstein's equations and the requirement of
very large computational resources. Numerical simulations are
currently only in their infancy~\cite{on96}.  Recently, Wilson and
Mathews~\cite{wm95} reported preliminary results obtained with a
relativistic numerical evolution code. Their dynamical calculations
suggest that the neutron stars may collapse to black holes prior to
merger.  They also find that, typically, binaries have too large a
total angular momentum to form a Kerr black hole immediately upon
merger (see also~\cite{eh95}). Their results are in disagreement with
predictions of Newtonian~\cite{lrs93} and PN calculations~\cite{l96},
which show that tidal fields stabilize neutron stars against radial
collapse.

In this Letter we report the first calculations in full relativity of
quasi-equilibrium, equal mass, neutron star binaries in synchronized
circular orbits. We numerically integrate a subset of the Einstein
equations, coupled to the equations of relativistic hydrodynamics, to
solve the initial value problem for binaries.  Such quasi-equilibrium
models provide initial data for future dynamical evolution
calculations. We construct quasi-equilibrium sequences of constant
rest mass configurations at varying separation. These sequences mimic
evolutionary sequences in which the stars undergo slow inspiral on
nearly circular orbits due to the emission of gravitational waves.  We
identify the innermost stable circular orbit (ISCO), its angular
velocity, and the maximum quasi-equilibrium mass of a neutron star in
a close binary.

In Newtonian gravity, strict equilibrium for two stars in
synchronized circular orbit exists.  Since this solution is
stationary, the hydrodynamical equations reduce to the Bernoulli
equation, which greatly simplifies the problem.  Because of the
emission of gravitational waves, a binary in general relativity cannot
be in strict equilibrium. However, outside the ISCO, the timescale for
orbital decay by radiation is much longer than the orbital period, so
that the binary can be considered to be in ``quasi-equilibrium''. This
fact allows us to neglect both gravitational waves and wave-induced
deviations from a circular orbit to very good approximation~\cite{se}.
Some of our approximations have been used and calibrated
elsewhere~\cite{wm89,cst96}, and a more detailed discussion will be
presented in a forthcoming paper~\cite{bcsst97}. Here we will briefly
outline our method and present some of our key results.

We attempt to minimize 
the gravitational wave content in the solution, in compliance
with physical expectations, by choosing the 3-metric to
be conformally flat~\cite{wm95,wm89}. In cartesian coordinates
the line element can then be written 
\begin{equation}
ds^2 = - \alpha^2 dt^2 + \Psi^4 \delta_{ij}(dx^i - \omega^i dt)
	(dx^j - \omega^j dt),
\end{equation}
where $\alpha$ is the lapse, $\omega^i$ the shift and $\Psi$ the
conformal factor. We satisfy the initial value equations of relativity
precisely. Our approximation lies in assuming that the metric will
remain conformally flat for all times. The extrinsic curvature
$K_{ij}$ then has to satisfy
\begin{equation} \label{k1}
\bar K^{ij} = - \frac{\Psi^6}{2\alpha} 
	\left( \nabla^i \omega^j + \nabla^j \omega^i
	- \frac{2}{3} \delta^{ij} \nabla_k \omega^k \right),
\end{equation}
where $\bar K^{ij} = \Psi^{10} K^{ij}$ and where we have also used the
maximal slicing condition $K^i_{~i} = 0$. Here $\nabla^i$ is
the flat space derivative operator in cartesian coordinates.

We assume that the matter obeys a polytropic equation of state
\begin{equation} \label{eos}
P = K \rho_0^{1 + 1/n},
\end{equation}
where $P$ is the pressure, $\rho_0$ the rest-mass density, $K$ the
polytropic constant, and $n$ the polytropic index. We assume that we
can neglect deviations from a strictly periodic circular orbit and
that the stars are corotating, which is equivalent to assuming that
the fluid four-velocity is proportional to a Killing vector. In this
case the matter equations can be integrated analytically, which yields
the relativistic Bernoulli equation
\begin{equation} \label{bernoulli}
q = \frac{1}{1+n} \left( \frac{1 + C}{\alpha (1 - v^2)^{1/2}} -1 \right),
\end{equation}
where $q = P/\rho_0$, $C$ is a constant of integration and $v$
is the proper velocity of the matter.

The Hamiltonian constraint can now be written
\begin{eqnarray} \label{ham}
\nabla^2 \Psi & = & -\frac{1}{8} \Psi^{-7} \bar K_{ij} \bar K^{ij} - \\[1mm]
& &	2 \pi \Psi^5 q^n \left( \frac{1+(1+n)q}{1-v^2} - q \right).
\nonumber
\end{eqnarray}
Requiring that the maximal slicing condition be maintained at all times,
we can use the time evolution equation for $K_{ij}$ to find an equation
for the lapse,
\begin{eqnarray} \label{lapse}
\nabla^2 \tilde \alpha & = & \frac{7}{8} \tilde \alpha \Psi^{-8}
 	\bar K_{ij} \bar K^{ij} + \\[1mm]
& &	2 \pi \tilde \alpha \Psi^4 q^n 
	\left( (1+(n+1)q) \frac{1+2v^2}{1-v^2} + 5q \right),
\nonumber
\end{eqnarray}
where $\tilde \alpha = \Psi \alpha$. The momentum constraint becomes
\begin{eqnarray} \label{g}
\nabla^2 \omega^i + \frac{1}{3} \nabla^i (\nabla_j \omega^j)
	& = & -2 \nabla_j (\tilde \alpha \Psi^{-7}) \bar K^{ij} - \\[1mm]
& &	16 \pi \Psi^4 q^n \frac{1+(1+n)q}{1-v^2}(\Omega\xi^i - \omega^i),
\nonumber
\end{eqnarray}
where $\Omega$ is the constant angular velocity and $\xi^i$ is a 
three-vector tangent to the matter velocity. With
the stars centered along the $z$-axis and orbiting
around the $y$-axis, we have $\xi^i = (z,0,-x)$. The last equation can
be simplified by writing $\omega^i = G^i - \frac{1}{4} \nabla^i B$.

Our approximations reduce the Einstein field equations to a set of
coupled, quasi-linear elliptic equations for the lapse, shift, and the
conformal factor (eqs.~(\ref{ham}--\ref{g})) which have to be solved
together with the matter equation~(\ref{bernoulli}). For boundary
conditions at large radius we impose asymptotic flatness.  Solving
these equations yields a valid solution to the initial value
(constraint) equations. Such a solution will also provide an
approximate instantaneous snapshot of a binary evolved according to
the full Einstein equations, prior to plunge.  In the Newtonian limit,
the above equations reduce to the coupled Poisson and Bernoulli
equations.

Our numerical implementation will be described in detail
in~\cite{bcsst97}. Since the stars have equal mass, it is sufficient
to work in one octant only.  We use a full approximation storage
multigrid scheme to solve the elliptic field
equations~(\ref{ham}--\ref{g}) for a given matter distribution. Once a
solution has been found, the matter can be updated
(eq.~(\ref{bernoulli})). This iteration can be repeated until
convergence is achieved to a desired accuracy. We have implemented
this algorithm in a parallel environment using DAGH
software~\cite{pb95}. Typical runs were computed on a grid of $64^3$
gridpoints. We adjusted the outer boundaries for each separation so
that the matter was always covered by 17 gridpoints along the
diameter.

We determine the rest (baryon) mass $M_0$, the total mass-energy (ADM
mass) $M$, as well as the angular momentum $J$, which refer to the
parameters of one individual star.  Note that physical dimensions
enter the problem only through the polytropic constant $K$
in~(\ref{eos}). It is therefore convenient to introduce the
dimensionless quantities $\bar \rho_0 = K^n \rho_0$, $\bar M_0 =
K^{-n/2} M_0$ and $\bar M = K^{-n/2} M$.

\begin{figure}
\epsfxsize=2.5in
\begin{center}
\leavevmode
\epsffile{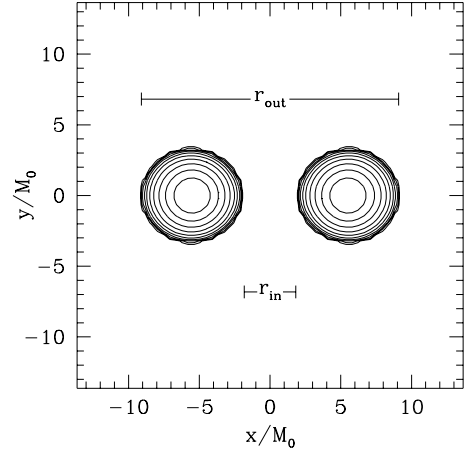}
\end{center}
\caption{Rest-density contours in the equatorial plane for a neutron 
star binary close to the ISCO.  Each star has a rest mass of $\bar M_0
= 0.178$, only slightly below the maximum mass at infinite separation,
$\bar M_0^{\rm max} = 0.180$.  The contours span
densities between the central density and 1\% of that value by
decreasing factors of 0.63.}
\end{figure}

In the following we will discuss results for $n = 1$.  A survey of
several different polytropic indices between 1 and 2.9 will be
presented in~\cite{bcsst97}. 

In Fig.~1 we show density profiles for highly relativistic neutron
stars of rest mass $\bar M_0 = 0.178$ close to the ISCO. The maximum
mass for such a star in isolation is $\bar M_0^{\rm max} =
0.180$. Note that the stars at the ISCO are only very slightly
distorted.  In the following we will define the ratio between the
inner and outer coordinate separation, $z_A = r_{in} / r_{out}$.

In Fig.~2 we plot the allowed rest mass versus the central density for
several different separations between $z_A = 0.3$ (roughly two stellar
radii apart) to $z_A = 0$ (touching).  As $z_A \rightarrow 1$ we
expect these curves to approach the spherical Oppenheimer-Volkoff (OV)
result, which we included as the dashed line in Fig.~2.  All our
graphs lie within 2\% of the OV curve, showing that the presence of a
companion star has only very little influence on the mass-density
equilibrium relationship.

\begin{figure}
\epsfxsize=2.5in
\begin{center}
\leavevmode
\epsffile{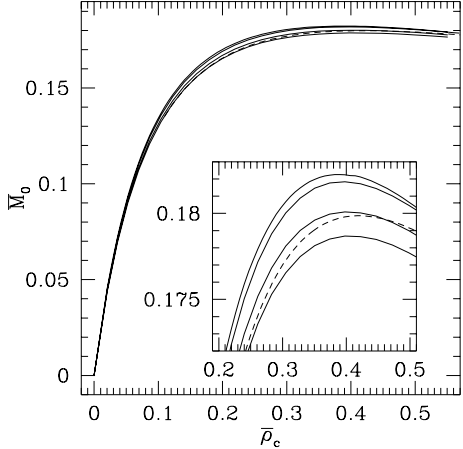}
\end{center}
\caption{Rest mass $\bar M_0$ versus central density $\bar\rho_{\rm c}$
for separations $z_A = 0.3$ (bottom solid line), 0.2, 0.1 and 0.0 (top line).
The dashed line is the Oppenheimer-Volkoff result. The insert is a blow-up
of the region around the maximum masses.}
\end{figure}

As we decrease the separation, the mass supported by a given $\bar
\rho_{\rm c}$ increases slightly. In particular, the maximum mass
increases from $\bar M_0^{\rm max} = 0.179$ for $z_A = 0.3$ to $\bar
M_0^{\rm max} = 0.182$ for touching stars. This trend clearly suggests
that {\em the maximum allowed mass of neutron stars in close binaries
is slightly larger than in isolation}. This increase is caused partly
by the rotation of the stars and partly by the tidal
fields~\cite{rot}.  Note, however, that we are only constructing {\em
quasi-equilibrium} configurations, which may or may not be dynamically
stable. For nonrotating, isolated stars the maximum mass configuration
in Fig.~2 marks the onset of radial instability. No general theorem can
be trivially applied to binary stars. If the results~\cite{wm95} are
correct (but see~\cite{l96,lrs93}), it is because the equilibrium
configurations are dynamically unstable, and not because the maximum
allowed mass decreases.  Fig.~2 also shows that keeping the rest mass
fixed, the central density slightly decreases as the stars approach
each other and become tidally deformed. Both effects are consistent
with simple PN predictions~\cite{l96,lrs97}.

\begin{figure}[t]
\epsfxsize=2.5in
\begin{center}
\leavevmode
\epsffile{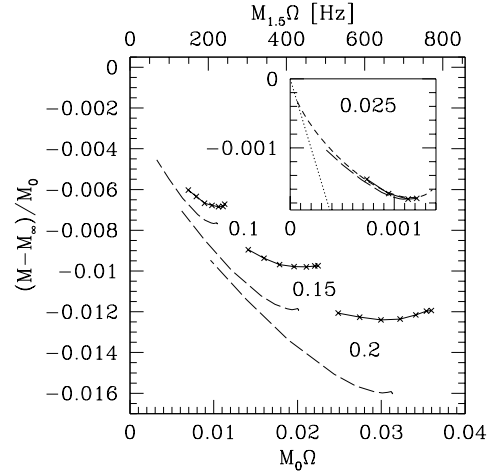}
\end{center}
\caption{The binding energy as a function of the angular velocity
for several different values of the rest mass $\bar M_0$.  The curves
are labeled by the compaction $(M/R)_{\infty}$ of the stars in
isolation at infinity. The maximum compaction for a stable, isolated,
nonrotating $n=1$ polytrope is $0.217$.  The upper label gives the
frequency for stars with a rest mass of $1.5 M_{\odot}$.  The dashed
curves are the corresponding results from a Newtonian version of our
code. In the insert we plot results for a nearly Newtonian
configuration with $(M/R)_{\infty} = 0.025$. Here we have also
included the results from a Newtonian ellipsoidal treatment~[12] (long
dashed line) and for two point particles (dotted line).}
\end{figure}

We construct sequences of constant rest mass $\bar M_0$, which
approximate evolutionary sequences up to the ISCO~\cite{ct}.  Note
that we have assumed the stars to be corotating. This may not be
realistic, since it would require excessive
viscosity~\cite{viscosity}. It is more likely that circulation of the
stars is conserved during inspiral, and the stars remain nearly
irrotational. Nevertheless, we expect that our sequences are a
reasonable approximation to the inspiral up to the ISCO and
correctly reveal the effects of nonlinear gravitation.

In Fig.~3 we show the binary binding energy versus angular velocity
for several different rest masses $\bar M_0$. As the stars approach,
both finite size effects and nonlinear gravitation play an
increasingly important role and cause, for stiff enough equations of
state, the binding energy to go through a minimum and increase again.
The location of the minimum marks the onset of a secular instability,
beyond which the binary can no longer maintain circular
equilibrium. It is expected that the dynamical instability, which
defines the true ISCO for plunge, occurs beyond, but close to, the
onset of the secular instability~\cite{lrs93}.

In the Newtonian regime our results agree very well with both a
Newtonian version of our code and results from an ellipsoidal
treatment of the binaries~\cite{lrs93} (see insert). For more
relativistic models, comparisons are made somewhat ambiguous by the
adopted choice of a parameter to characterize the
sequence. Identifying the member at infinity by its value of $M$ (or
$M_0$) versus $(M/R)_{\infty}$ (or $(M_0/R)_{\infty}$) leads to
different Newtonian models and binding energy curves, and the
differences increase as the stars become more compact. In Fig.~3 we
choose $(M_0/R)_{\infty}$ and find that the ISCO frequencies agree
closely with the Newtonian values, but the binding energies differ as
the compaction increases~\cite{pnc}.

We summarize these results in Table~1, where we also tabulate the
dimensionless total angular momentum $J_{tot}/M_{tot}^2 = J/2M^2$ at
the ISCO. Note that for high enough rest masses this value drops below
unity, so that these two stars could plunge and form a Kerr black hole
without having to radiate additional angular momentum. Because the
orbit will decay rapidly inside the ISCO, its presence will leave a
measurable imprint on the emitted gravitational wave form. Measuring
$\Omega_{ISCO}$ may be the crucial ingredient in determining the
radius of the star, assuming that the mass has been determined during
the prior inspiral phase~\cite{t95}.

It is a pleasure to thank Manish Parashar for his help with the
implementation of DAGH, and Andrew Abrahams, James Lombardi and Fred
Rasio for several helpful discussions. This work was supported by NSF
Grant AST 96-18524 and NASA Grant NAG 5-3420 at Illinois, NSF Grant
PHY 94-08378 at Cornell, and by the NSF Binary Black Hole Grand
Challenge Grant Nos.~NSF PHY 93-18152/ASC 93-18152 (ARPA
supplemented).

\begin{table}
\begin{center}
\begin{tabular}{ccccc}
$\bar M_0$  & $\bar M_{\infty}$ & $(M/R)_{\infty}$ & $M_0 \Omega_{ISCO}$ &
	$(J_{tot}/M_{tot}^2)_{ISCO}$  \\
\tableline
0.112	& 0.106		& 0.1	& 0.01	& 1.22 \\
0.134	& 0.126		& 0.125	& 0.015	& 1.12 \\
0.153	& 0.142		& 0.15	& 0.02	& 1.05 \\
0.169	& 0.155		& 0.175	& 0.025	& 1.00 \\
0.1781 	& 0.1623      	& 0.2	& 0.03	& 0.97 
\end{tabular}
\end{center}
\caption{Numerical values for sequences of constant rest mass $\bar M_0$
and polytropic index $n=1$.  We tabulate the total energy $\bar
M_{\infty}$ and compaction $(M/R)_{\infty}$ each star would have in
isolation at infinity as well as the angular velocity $M_0 \Omega$ and
the angular momentum $J/M^2$ at the ISCO.}
\end{table}


\begin{references}
\bibitem{se} Note that a similar approximation is often used in 
	stellar evolution calculations. There the relevant evolution
	timescales are the nuclear or Kelvin-Helmholtz timescales, while 
	the stars maintain (quasi) hydrostatic equilibrium 
	on a dynamical timescale.
\end{references}
\end{document}